\documentclass[%
showpacs,amssymb, amsmath,nobibnotes,superscriptaddress ,aps, prd]{revtex4}

\usepackage{graphicx}

\newcommand{\be}{\begin{equation}}
\newcommand{\ee}{\end{equation}}
\newcommand{\figwidth}{4in}

\begin{document}

\preprint{Liverpool Preprint: 617}

\title{An estimate of the flavour singlet contributions to the hyperfine splitting in 
charmonium.}
\author{C.~McNeile}
\email{mcneile@amtp.liv.ac.uk}
\author{C.~Michael}
\collaboration{UKQCD Collaboration}
\affiliation{
Theoretical Physics Division, Dept. of Mathematical Sciences, 
          University of Liverpool, Liverpool L69 3BX, UK}


\begin{abstract}
We explore the  splitting between  flavour  singlet
and non-singlet mesons in charmonium. This has
implications for the hyperfine splitting in  charmonium. 
\end{abstract}

\maketitle


\section{Introduction and motivation} \label{se:hyperFINE}

In the past year there have been many new interesting experimental
discoveries in meson spectroscopy with heavy
quarks~\cite{Stone:2003pc}.  
One of the goals of the experimental program at CLEO-c~\cite{Bianco:2003vb} 
is to refine our experimental knowledge of the heavy hadron spectrum.
This new data helps to validate methods, such as lattice gauge theory,
that solve 
QCD non-perturbatively.

On the theoretical side,
it has been claimed that there has been much progress in unquenched
lattice QCD calculations~\cite{Davies:2003ik}. 
These new lattice QCD calculations
that use improved staggered quarks have passed some important
experimental consistency checks~\cite{Davies:2003ik}.
However, the one place where the agreement
between lattice QCD and  and experiment is still poor is the mass
splitting between the  $J/\psi$ and $\eta_c$~\cite{DiPierro:2002ta}. 
The masses of these two mesons can usually be computed with the smallest
statistical errors. Also, as these masses are independent of light
valence quarks, this splitting does not depend on a large extrapolation
in the valence quark mass. It does of course depend on an extrapolation
in the  sea quark mass. 

The experimental value for the mass splitting between the $J/\psi$ and
$\eta_c$ is 116 MeV.   Some of the older lattice 
calculations~\cite{Allton:1992zy,El-Khadra:1993ir,Davies:1995db,
Trottier:1997ce,Shakespeare:1998dt,Mathur:2002ce,
Klassen:1998fh,Chen:2000ej,Okamoto:2001jb}
that computed this splitting have been reviewed
recently~\cite{McNeile:2002uy}.

When using a clover improved fermion formalism on the lattice, the
hyperfine splitting is sensitive to the coefficient $c_{SW}$ 
(of a term in the fermion operator that helps reduce lattice spacing
dependence on physical quantities)
at nonzero
lattice spacing, but the hyperfine splitting should be independent of
the $c_{SW}$ as the continuum limit is taken, because the clover term is
an irrelevant operator. Recently the QCD-TARO
collaboration~\cite{Choe:2001yg} have studied the charmonium spectrum in
quenched QCD using the clover action at a smaller lattice
spacing ($a^{-1} \sim$ 5 GeV) than previously used. They obtained
a hyperfine splitting of $77(2)(6)$ MeV 
in the 
continuum limit.

There has not been much work on the charmonium spectrum from unquenched
lattice QCD calculations. There are arguments  based on potential
models, that suggest the hyperfine splitting in charmonium is sensitive
to the presence of light sea
quarks~\cite{El-Khadra:1993ir,Bernard:2000gd}. El-Khadra et
al.~\cite{El-Khadra:2000zs}  did look at the charmonium spectrum on 
(unimproved) staggered gauge configurations  ($m_{\pi}/m_{\rho}$ was 0.6
and the lattice spacing was  $a^{-1} \sim $ 0.99(4) GeV.) from the MILC
collaboration.  No significant increase in the hyperfine splitting was
reported. Stewart and Koniuk~\cite{Stewart:2000ev} studied the
charmonium spectrum using NRQCD on unquenched (unimproved) staggered 
gauge configurations ($m_{\pi}/m_{\rho} \sim$ 0.45  and $a \sim$ 0.16
fm). Any signal for the effect of unquenching was hidden beneath the
systematic uncertainties in  using the NRQCD formalism
for charmonium.

The work by Davies et al.~\cite{Davies:2003ik,Bernard:2001av}  found that
the correct ratio was produced for the (P-S)/(2S-1S) mass splittings for
$\bar{b}b $ using unquenched calculations with improved staggered fermions.
 However, the hyperfine splitting in charmonium is still 
incorrect~\cite{DiPierro:2002ta,diPierro:2003bu,Gottlieb:2003bt} at  
$97 \pm 2$ MeV.  The authors claim that the discrepancy may be caused
by the clover
coefficient only being used to tree level in tadpole improved perturbation
theory. These calculations do, however, get the hyperfine  splitting,
the so-called J parameter~\cite{Lacock:1995tq}, in the light quark
sector correct~\cite{Bernard:2001av}.

\begin{figure}[th]
\includegraphics[width=\figwidth,clip]{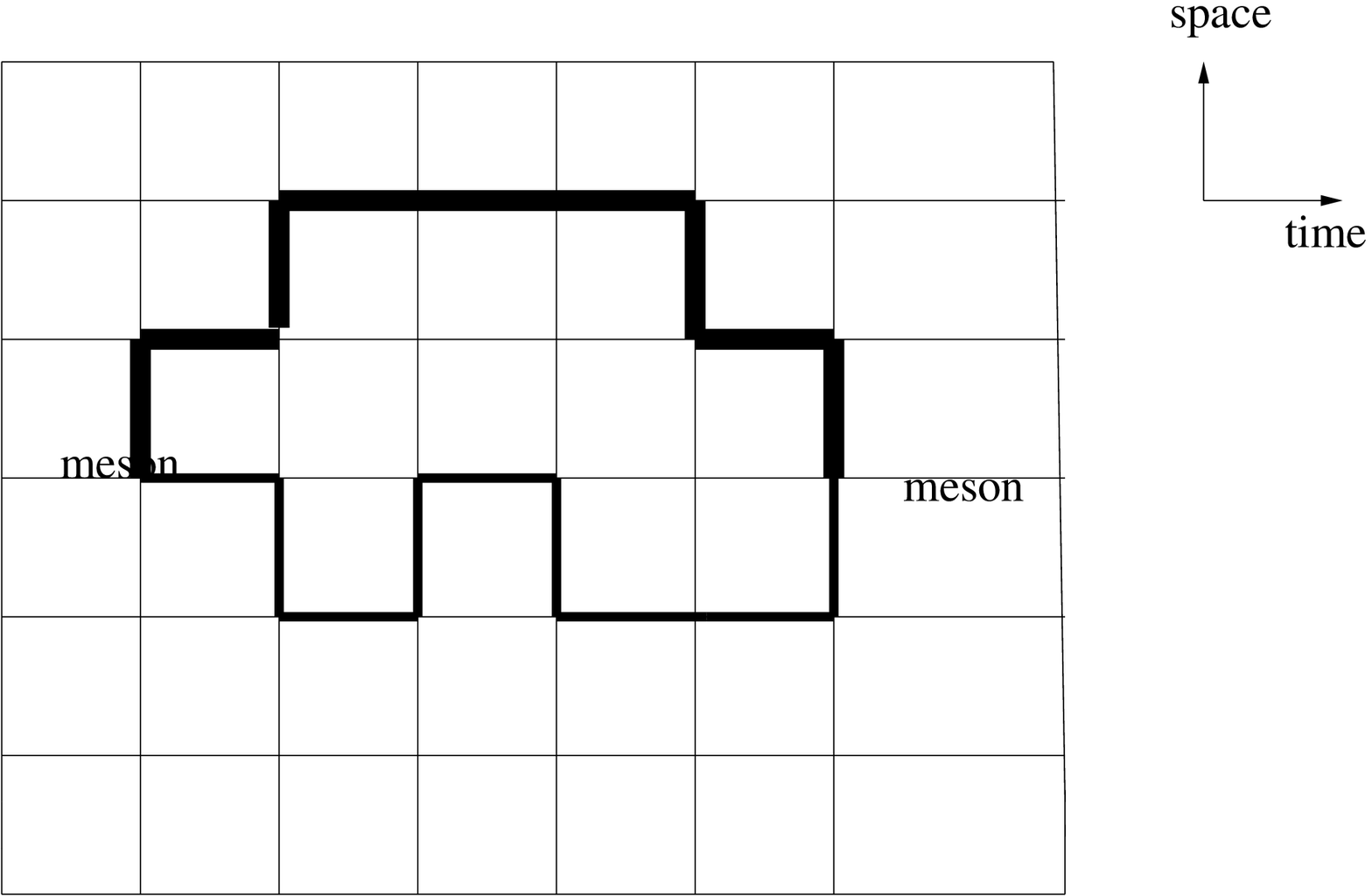}
\caption{Connected loop contribution to the propagator}
\label{fg:conLOOPS}
\end{figure}

\begin{figure}[th]
\includegraphics[width=\figwidth,clip]{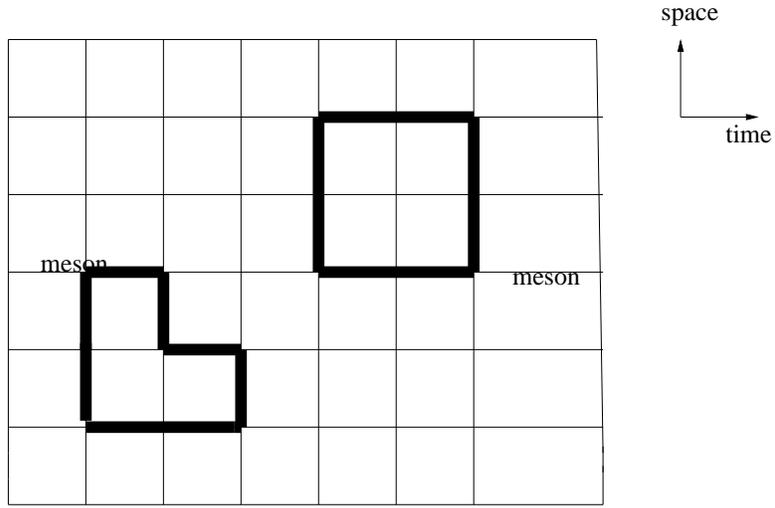}
\caption{Disconnected loop contribution to the propagator}
\label{fg:disLOOPS}
\end{figure}

There is another possible reason that the hyperfine mass splitting
between the $J/\psi$ and $\eta_c$ is smaller than experiment in current
lattice evaluations.  This has been also discussed recently by
QCD-TARO~\cite{Choe:2003wx,McNeile:2002uy,Bali:2003tp}.  All lattice calculations
have computed the non-singlet correlator (see figure~\ref{fg:conLOOPS}).
However, charmonium interpolating operators are actually singlet
($\overline{c}\, \Gamma c$), so the Wick contractions contain bubble
diagrams (see figure~\ref{fg:disLOOPS}).  The bubble diagrams are OZI
suppressed so should be small.  However, this argument will fail if
there is additional non-perturbative physics. For light
mesons~\cite{McNeile:2001cr}, it has been found that the effects of the
bubbles can be large for the pseudoscalar and scalar mesons where the
additional physics is the anomaly and the $0^{++}$ glueball
respectively, but not for other channels.  Essentially, the disconnected
loops are large if there is additional interesting physics such as
glueballs, the anomaly, or instantons. It is possible to explore this
non-perturbatively from the lattice and  we discuss these mechanisms in
section~\ref{sect:discuss}.

It is interesting to compare the hyperfine splitting  for $D$ mesons with
that in charmonium, as there is no contribution from the bubble diagrams
for the  $D$ mesons. In table~\ref{tab:hyperfine} we have collected some
results for  the $D^\star-D$ mass splitting from quenched QCD.
 The agreement between experiment and lattice is pretty good for the
mass splitting between the $D^\star$ and $D$.  The differences could be
explained by the ambiguity in determining the lattice spacing  in
quenched QCD. As noted recently by di Pierro et
al.~\cite{diPierro:2003bu},  the hyperfine splitting in the D system is
first order in the clover coefficient, but the  hyperfine splitting in
charmonium is second order in the  clover coefficient. Hence, the
hyperfine splitting in charmonium  may be more sensitive to the clover
coefficient. The differences  in the hyperfine splitting may potentially
be due to remaining errors in the determination of $c_{SW}$.

\begin{table}[tb]
\begin{center}
\begin{tabular}{|c|c|c|} \hline
Group & Method &  $M_{D^{\star}}-M_{D}$ MeV \\  \hline
Boyle~\cite{Boyle:1997aq} & clover &  124(8)(15) \\
Boyle~\cite{Boyle:1998rk} & $\beta$=6.0 tadpole clover &  106(8) \\
Hein et al.~\cite{Hein:2000qu} & NRQCD  $\beta = 5.7$ & 
$110^{+3+22}_{-0-0}(3)(6)(5)$ \\
UKQCD~\cite{Bowler:2000xw} & NP clover  $\beta = 6.2$ & 
$130^{+6+15}_{-6-35}$ \\
UKQCD~\cite{Flynn:2003vz} & NP clover  $\beta = 6.2$ & 
$127(14)(1)(3)$
\\ \hline 
PDG~\cite{Hagiwara:2002fs} & Experiment  & 142.12(7) 
\\ \hline 
  \end{tabular}
\end{center}
  \caption{
Collection of hyperfine splittings between the
$D$ and \protect{$D^{\star}$ mesons}.
}
\label{tab:hyperfine}
\end{table}

 It is clearly not sufficient to just assume that the OZI-violating 
disconnected contributions to charmonium states are negligible,
particularly as they  are responsible for the decay width of $\eta_c$ of
some tens of MeV.  In order to clarify these issues, we explore from first
principles the  importance of disconnected contributions to  the
charmonium hyperfine splitting.

\section{Singlet correlators}

In lattice studies it is possible to measure separately  the 
non-singlet contribution given by connected correlation $C(t)$ (see
figure~\ref{fg:conLOOPS}) and  the flavour singlet contribution which
has an additional  disconnected correlation  $D(t)$ (see
figure~\ref{fg:disLOOPS}). 
 Previous lattice studies have been made of the  light pseudoscalar
mesons~\cite{Kuramashi:1994aj,AliKhan:1999zi,Bardeen:2000cz,
Struckmann:2000bt,McNeile:2000hf} and  scalar mesons
~\cite{Lee:1998gi,Lee:1998yd,Lee:1999kv,McNeile:2000xx,Hart:2002sp}. 
For a discussion including some results for vector and
axial mesons, see~\cite{Isgur:2000ts}.

 In the flavour singlet case there is an additional
disconnected correlation  $D(t)$ to be evaluated.  This  correlation can
be written in the form  
 \begin{equation}
D(t)= N_f r_4 r_5 \langle L(0) L^*(t)  \rangle
\label{eq:DEXPR}
 \end{equation}
where the disconnected loop
 \begin{equation}
L(t) =  {\rm Tr} \, { \Gamma M^{-1} } 
\label{eq:loopEXPR}
 \end{equation}
 with $M^{-1}$ the quark propagator  and the sum in the trace is over
colour, Dirac and spatial indices at time $t$. Here we assume that the
hadron  under consideration is created by $\bar{q} \Gamma q$ and so the
factor of  $r_4$ arises from reflection positivity (i.e. $\gamma_4
\Gamma^{\dag}=r_4 \Gamma \gamma_4$).
 The factor of $r_5$ arises  since the Wilson-Dirac fermion matrix $M$ is
$\gamma_5$ hermitian and hence $L$ is real/ imaginary as $\gamma_5
\Gamma = r_5 \Gamma \gamma_5$ with $r_5=\pm 1$. Since at  $t=0$ we have that
$L(0) L^*(0) > 0$,  the disconnected correlation  $D(0)$ has sign $r_4
r_5$. 

At large t where ground state contributions dominate  we have
 \be  C(t) =  c e^{-m_1 t}  \ee
 and
 \be  C(t)+D(t) = d e^{-m_0 t}  \ee
 where  $m_0$ is the flavour singlet mass  and $m_1$ the flavour
non-singlet mass. Here we are ignoring   lighter
singlet pseudoscalar states (with no charm content, such as $\eta$)
which contribute  with very small amplitude to $C+D$.   If the same
meson
creation and destruction operators are used for the  study of  both
correlations, with quarks degenerate in   mass,  $d$ and $c$ have the
same sign.

 Then by a study of  $D/C$ which is given  by 
 \begin{equation} 
D(t)/C(t)= (d/c) e^{(m_1-m_0) t} -1 
\label{dbyc.eq}
 \end{equation}
 the mass splitting between flavour singlet and non-singlet can be
explored. Although it might be thought that $d=c$, we have shown 
previously~\cite{McNeile:2000xx} that this is  not necessarily the case,
and indeed sign changes in $D/C$ versus $t$  can be required.
  So, in summary,  the slope (increase/decrease) of $D/C$ on a lattice
can determine the  sign and magnitude of $m_1-m_0$. For charmonium it is
correct to use $N_f=1$ in  equation~\ref{eq:DEXPR} since only one
flavour  of quark can contribute to the loops. In our comparisons using
lighter  quarks for the loops, we will also use $N_f=1$. A complete
analysis of the  light two flavour correlator will appear
elsewhere~\cite{Hart:2002sp}. 

Since, as we shall see, the disconnected contributions are poorly
determined  as $t$ increases, it is advantageous to remove excited
state contributions as far as possible. One technique, pioneered by Neff
et al.~\cite{Neff:2001zr}, is  to study the ratio of $D/C$ using  the
ground state contribution to the connected correlator $C$ from a fit.
This will be appropriate if the  disconnected contribution $D$ has only
small excited state contributions, as  does seem to be the case.
 \begin{equation} 
D(t)/C(t)_{fit} = (d/c) e^{(m_1-m_0) t} -1 
\label{SESAMdbyc.eq}
 \end{equation}

\section{Lattice Methodology}

We use dynamical fermion configurations with  $N_f=2$ from
UKQCD~\cite{Allton:2001sk}. The sea quarks correspond to $\kappa=0.135$
with a  non-perturbative improved clover formalism. The volume was
$16^{3}32$. This data set has a scale set by~\cite{Allton:2001sk}  
$r_0/a=4.754(40)(+2-90)$ and pseudoscalar meson to vector meson mass
ratio of $m_{P}/m_{V}$ = 0.70. Using the value $r_0=0.525$
fm then gives $a^{-1}=1790$ MeV while the meson mass ratio implies that
the sea quarks have masses  close to that of a strange quark.  We have
already reported the spectrum of the charm-strange mesons and 
preliminary results for the  mass of the charm quark on this data 
set~\cite{Dougall:2003hv,Dougall:2003mx}.

Local and spatially-fuzzed operators~\cite{Lacock:1995qx} are used for
meson creation  (with two fuzzed links in a spatially symmetric
orientation with 5 iterations of fuzzing with  coefficient given by
2.5*Straight + Sum of staples).  Thus we evaluate a   $2 \times 2$
matrix of local and fuzzed correlators~\cite{Lacock:1995qx}.  Mesons
created by all independent products of gamma matrices are evaluated.

We measured the connected and  disconnected correlations on 201
configurations of size $16^3 32$ separated by 40 trajectories for three
heavy $\kappa$ values: 0.113, 0.119, 0.125.  This data set was used 
to estimate the $\kappa$ value for the mass of the charm quark.
Our preliminary
estimate for the $\kappa$ value at the charm quark mass was close to 0.119.
As the aim of this study was to look for the singlet contribution to the
charmonium correlators, we did additional runs at $\kappa = 0.119$. At
$\kappa=0.119$, we computed connected and disconnected correlators
separated by 10 sweeps, hence the ensemble size was 788. The correlators
were then blocked with a block size of 40 sweeps. At $\kappa = 0.119$,
all the  results reported here are from the higher statistics run.

For the evaluation of the disconnected correlators, we use 100
stochastic noise sources with the two source trick described
in~\cite{McNeile:2000xx}.  We use sources at every site on the lattice
and determine the momentum-zero correlations from them. The connected
correlators are obtained by explicit inversion from a source (local or
fuzzed)~\cite{Allton:2001sk}. 

\section{Analysis of the connected correlators}

The lattice spacing used in this data set is large relative to the mass
of the charm quark, hence lattice spacing errors are a potential
concern.  In quenched QCD it is computationally possible to use finer
lattice spacings, so lattice spacing errors can be controlled by ``brute
force''. The high cost of reducing the lattice spacing in unquenched
calculations means that a brute force approach will not be feasible for
many years with this type of fermion action. Hence we chose to investigate
the heavy quark formalism developed by the Fermilab
group~\cite{El-Khadra:1997mp}.

The lattice artifacts modify the 
dispersion~\cite{Bernard:1994zh,El-Khadra:1997mp} relation:
 \begin{equation}
E^2 = M_1^2 + \frac{M_1}{M_2}p^2 + O(p^4)
\label{eq:FNALdisp}
 \end{equation}
 where $M_1$ is known as the ``rest mass'' and  $M_2$ is the kinetic
mass (since $E = M_1 +p^2/(2M_2) + O(p^4)$). 
In the FNAL lattice heavy quark formalism~\cite{El-Khadra:1997mp},
the rest mass is affected 
by lattice artifacts, but the $M_2$ mass is the one that controls the
dynamics of the states. The quality of the disconnected data precluded
us obtaining any useful information from  the disconnected data with
non-zero momentum.
A definition of the kinetic mass ($M_2$) in terms of the 
energy $E$ of the meson is:
 \begin{equation}
\frac{1}{M_2} =2 \frac{\partial E}{\partial p^2}_{\mid p=0}
\end{equation}

There are a number of different ways to define the 
quark masses on the lattice.
The vector definition of the quark mass is
 \begin{equation}
m_{v} = \frac{1}{2} ( \frac{1}{\kappa} - \frac{1}{\kappa_{crit}} )
 \label{eq:vwi}
\end{equation}
where $\kappa_{crit}$ is the value of $\kappa$ where the pion mass
vanishes.
In tree level perturbation 
theory~\cite{El-Khadra:1997mp}, the kinetic definition of the 
the quark mass $m_2$ is related the vector definition of the quark
mass ($m_v$).
 \begin{equation}
m_1 = \ln (1 + m_v)
 \end{equation}
 \begin{equation}
\frac{1}{m_2} = \frac{2}{m_v(2 + m_v)}  +\frac{1}{1 + m_v} 
 \end{equation}
In the ALPHA
formulation~\cite{Luscher:1996sc} the  vector definition of the  quark
mass is  O(a) improved using 
 \begin{equation} 
 \hat{m}_v = m_v ( 1 + b_m m_v) 
 \end{equation}
 where the value of $b_m$ from perturbation theory is 
 \begin{equation}
 b_m = -\frac{1}{2} - 0.0962 g^2.
 \end{equation}
 There are different ways of including the $b_m$ term in the
calculations. The merits of the different ways are discussed in the
UKQCD paper~\cite{Bowler:2000xw} on heavy-light decay constants.
The tadpole improved expressions for the FNAL quark
masses are obtained by replacing $m_v$ with $m_v / u_0$, 
where $u_0 = 1/(8\kappa_{crit})$.

To find the value of $\kappa$ for the charm quark,
we interpolate the spin averaged heavy meson mass 
 \begin{equation}
M_{Sav} = \frac{1}{4} ( 3 M_{V} + M_{PS} )
\label{eq:spinAvdefn}
 \end{equation}
 linearly with the vector definition of the quark mass 
 to the experimental value at 3068 MeV. 
For the data set with correlators separated by 40 sweeps,
we fitted a two exponential factorising fit model to the 2 by 2
smearing matrix.
From that, using $M_1$, we obtain the kappa
value of 0.116 for the  charm quark. We comment later on the 
consequences of using other definitions of the mass.

\begin{figure}[th]
\includegraphics[angle=-90,width=\figwidth,clip]{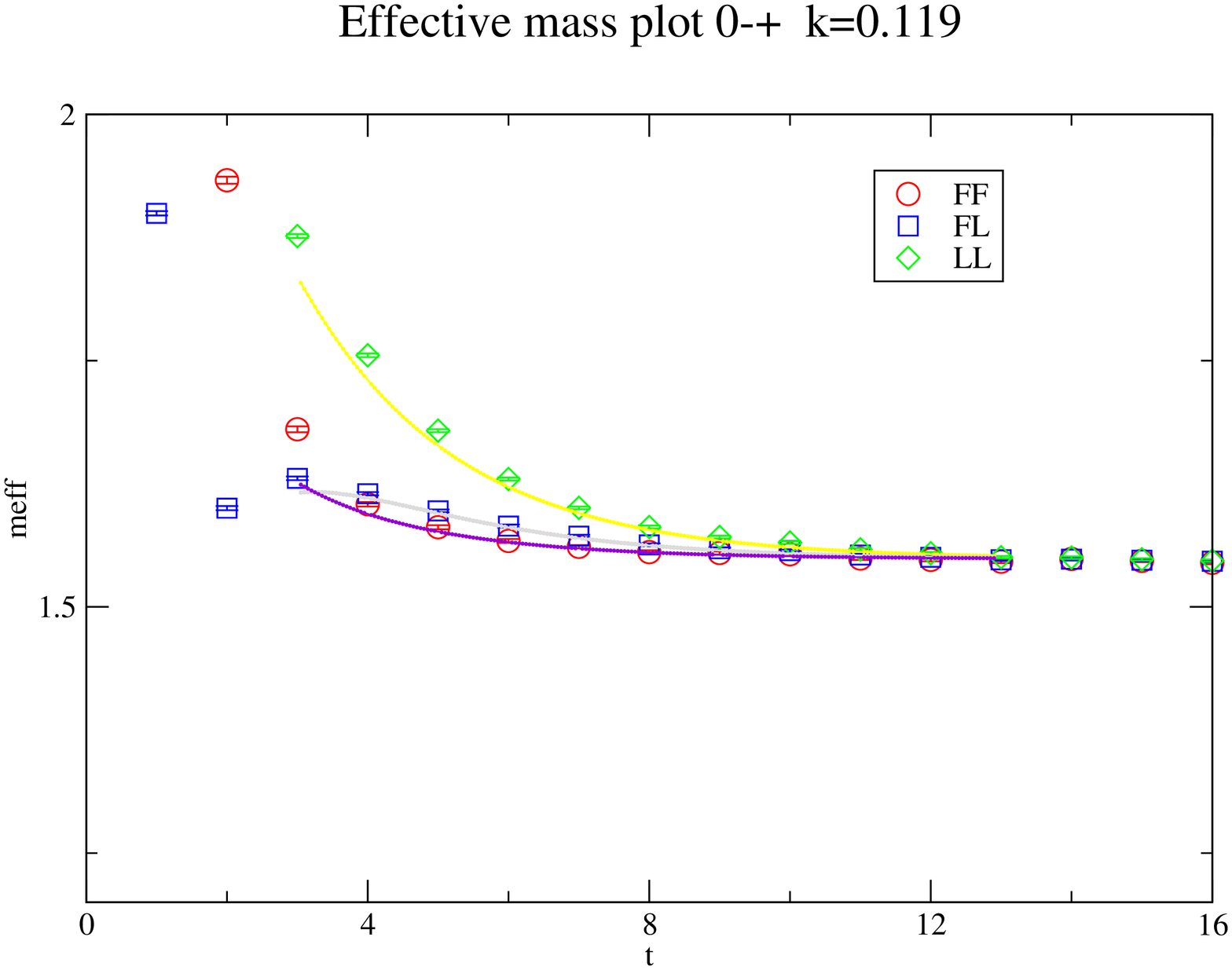}
\caption{Effective mass plot for the pseudoscalar at $\kappa$=0.119}
\label{fg:meffPS}
\end{figure}

\begin{figure}[th]
\includegraphics[angle=-90,width=\figwidth,clip]{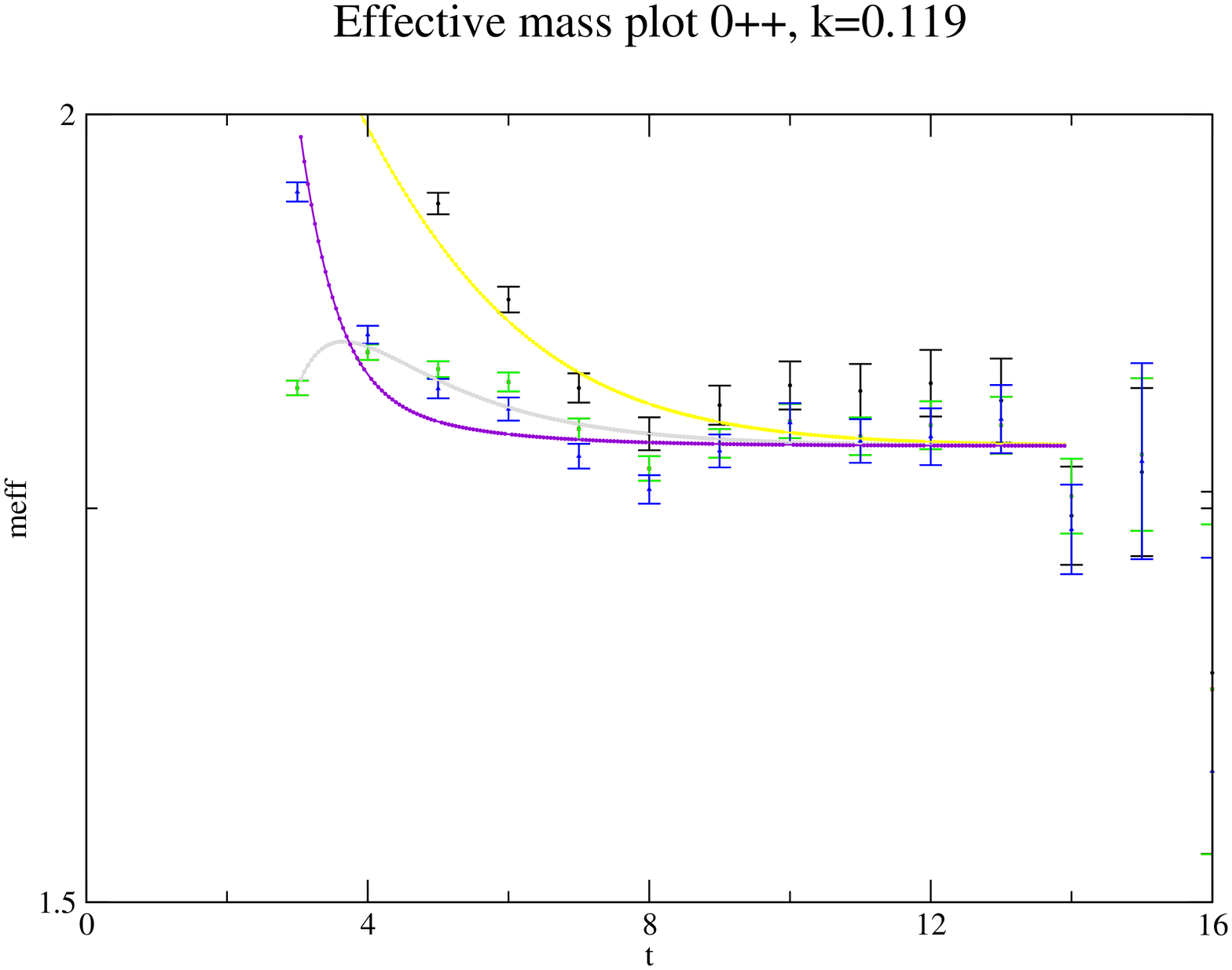}
\caption{Effective mass plot for the $0^{++}$ at $\kappa$=0.119}
\label{fg:meff0PP}
\end{figure}

\begin{figure}[th]
\includegraphics[angle=-90,width=\figwidth,clip]{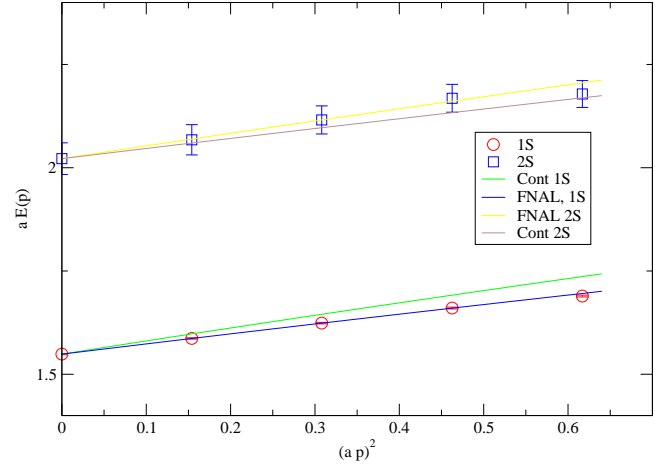}
\caption{Dispersion relation for the pseudoscalar at  $\kappa$=0.119}
\label{fg:DispPION}
\end{figure}

\begin{figure}[th]
\includegraphics[angle=-90,width=\figwidth,clip]{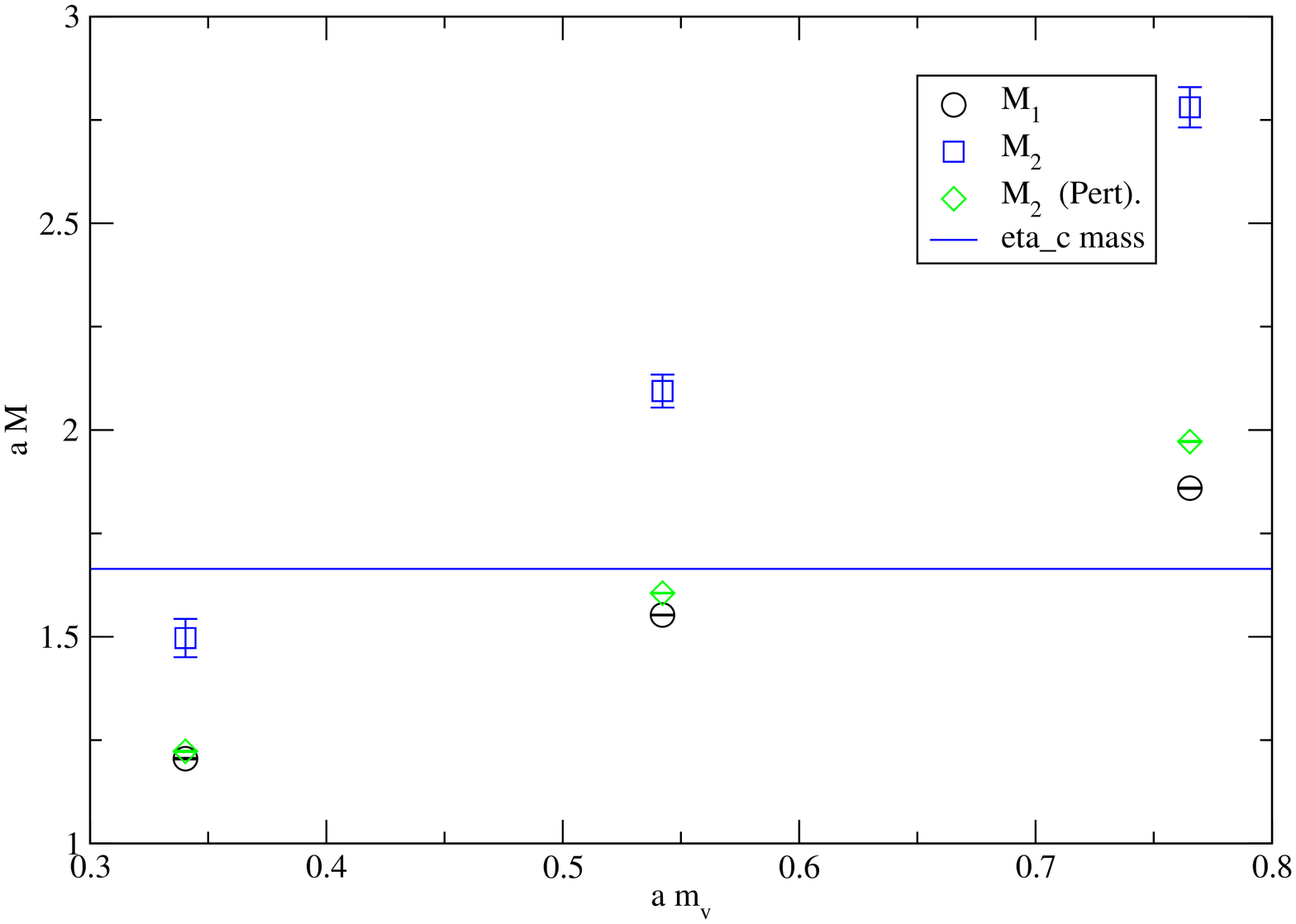}
\caption{Different definitions of the pseudoscalar meson mass versus the vector 
quark mass, in lattice units. This shows that the central quark mass 
value (from $\kappa=0.119$) is close to the experimental mass using $M_1$. 
The perturbative expressions are from equation~\ref{eq:PertMtwo} }
\label{fg:massDefn}
\end{figure}

We used ``factorising fits'' 
with three exponentials to fit the two by two matrix 
of smeared correlators for the higher statistics data set.
The effective mass plots for the pseudoscalar and 
scalar channels are in figures~\ref{fg:meffPS} and~\ref{fg:meff0PP}.
 The dispersion relation of the heavy-heavy pseudoscalar channel is
plotted in figure~\ref{fg:DispPION}. At the lattice spacing for this
calculation, the kinetic and rest masses differ by a significant amount.
In figure~\ref{fg:massDefn} we plot the various definitions of the meson
mass as a function of the  quark mass (defined from the vector Ward
identity, eq.~\ref{eq:vwi}). In quenched QCD at  finer lattice spacings
(0.07 fm) UKQCD~\cite{Bowler:2000xw} has shown that the $M_1$ and $M_2$
masses essentially agree. The high computational cost of reducing the
lattice spacing forces us to remain in a region where  $M_1$ and $M_2$
still differ  and we determine both masses. We fit
equation~\ref{eq:FNALdisp} to the  dispersion relation to calculate the
$M_2$ mass. The results for the various  masses are in
table~\ref{tab:connData}.

\begin{table}
\begin{center}
\begin{tabular}{|l|l|l|l|l|l|} \hline
Particle &  region & $\chi^2/dof$ & $a M_1$  & 
  $a M_2$ & $\frac{M_1}{M_2}$ \\  \hline
$0^{-+}$  & 3 - 13 & 2.7/24  &    1.549(1)  &  2.01(2)  &   0.772(8)  \\
$0^{-+}`$ & 3 - 13 & 2.7/24  &    2.02(3)   &  1.6(2)   &   1.3(2)    \\
$1^{--}$  & 3 - 13 & 2.3/24  &    1.593(2)  &  2.06(2)  &   0.772(8)  \\
$1^{--}`$ & 3 - 13 & 2.3/24  &    2.09(3)   &  1.6(2)   &   1.3(2)    \\
$0^{++}$  & 3 - 14 & 20/27   &    1.790(6)  &  2.5(2)   &   0.71(5)   \\
$0^{++}`$ & 3 - 14 & 20/27   &    2.38(5)   &   -       &     -       \\
$1^{+-}$  & 3 - 13 & 8.8/24  &    1.805(9)  &   -       &     -       \\
$1^{+-}`$ & 3 - 13 & 8.8/24  &    2.31(5)   &   -       &     -       \\
$1^{++}$  & 3 - 13 & 16.4/24 &    1.816(7)  &   -       &     -       \\
$1^{++}`$ & 3 - 13 & 16.4/24 &    2.39(5)   &   -       &     -       \\ 
\hline
\end{tabular}
\end{center}
\caption{Results for the masses from the fits to the connected data
at $\kappa = 0.119$. The masses of the ground and first excited
state are shown. The $M_1$ and $M_2$ masses are the meson
masses from the Fermilab formalism 
(from equation~\ref{eq:FNALdisp}). The fit regions and $\chi^2/dof$
are from the momentum zero fits.
 }
\label{tab:connData}
 \end{table}

The correlators with non-zero momentum are noisier  than those with zero
momentum. There are perturbative  expressions for the kinetic meson mass in
terms of the rest mass of the meson mass. At tree level
 \begin{equation}
M_2^{PT} = M_1 + (m_2 - m_1).
\label{eq:PertMtwo}
 \end{equation}
Our results for the $M_1$, $M_2$, and $M_2^{PT}$ are
plotted against the vector definition of the quark mass
in figure~\ref{fg:massDefn}. The tree level perturbative
expression does not fit the numerical data very well.

 Table~\ref{tab:connDataSplitt} contains the mass splittings from the
connected correlators. The results for the $M_2$ mass for the excited
states are very peculiar: we do see a linear behaviour of $E^2$ versus
$p^2$  so that $M_2$ can be extracted, but for the excited states it is
{\em smaller} than for the ground state. This illustrates the limitation
 of interpreting $M_2$ as the meson mass and implies that a more
sophisticated  treatment is needed to deal with heavy
quarks~\cite{Aoki:2001ra}. The JLQCD collaboration also
argue  that the kinetic mass is not necessarily superior 
to the pole mass~\cite{Aoki:2001ra,Aoki:1997xe}.

 \begin{table}
\begin{center}
\begin{tabular}{|l|ll|ll|l|} \hline
Splitting &  a $\Delta M_1$ & $\Delta M_1$ MeV &
a $\Delta M_2$ & $\Delta M_2$ MeV & Expt. MeV
\\  \hline
$1^{--} - 0^{-+}$ &
0.0446(6)     &    80(1) &
0.06(1)       &    105(19) &
116 \\
$1^{--}(2S) - 0^{-+}(2S)$ & 
0.07(1)       &    126(24) &
-0.01(10)      &   -18(189) &
32 \\
$0^{++} - S_{av}$   &
0.207(6) & 371(11)  &
0.5(2)   & 864(330) & 
348 \\
$0^{-+}(2S) - S_{av}$  & 0.44(3) & 787(59) &
- & - & 587  \\
\hline
\end{tabular}
\end{center}
\caption{
 Mass splittings from this data set. $S_{av}$ is the spin averaged  mass
(see equation~\ref{eq:spinAvdefn}). The experimental numbers come from
the particle data group.
 }
\label{tab:connDataSplitt}
 \end{table}

There is a lot of experimental interest in the mass  of the $\eta_c(2S)$
meson~\cite{Skwarnicki:2003wn}.
Until recently the mass of the $\eta_c(2S)$ determined from the 
Crystal Ball
collaboration~\cite{Edwards:1982mq} was larger 
than the predictions from potential models~\cite{Martin:1982nw}. 
However, the new results for the mass of the 
$\eta_c(2S)$ 
from CLEO, BABAR, and BELLE are in much better
agreement with potential models~\cite{Skwarnicki:2003wn}.
In table~\ref{tab:hyperfineTwoS} we collect the results from some
quenched lattice calculations and some recent experiments. 
Our result for the first excited hyperfine splittings is probably
effected by lattice spacing errors. We discuss the effect of 
glueballs on the $\eta_c(2S)$ meson in section~\ref{sect:discuss}.


\begin{table}[tb]
\begin{center}
\begin{tabular}{|c|c|c|} \hline
Group & Method &  $M_{\psi(2S)} - M_{\eta_c(2S)}$ MeV \\  \hline
Columbia~\cite{Chen:2000ej} & anisotropic, lattice  &  75(44) \\
CP-PACS~\cite{Okamoto:2001jb} & anisotropic, lattice & 26(17)  \\
\hline 
PDG~\cite{Hagiwara:2002fs}/Belle~\cite{Choi:2002na} & Particle data table
& $32 \pm 6 \pm 8$ \\
CLEO~\cite{CLEO:2003bk}    & Experiment  & $43 \pm 4 \pm 4$  \\
BABAR~\cite{Aubert:2003pt}  & Experiment  & $55 \pm 4 $  \\
Crystal Ball~\cite{Edwards:1982mq} & Experiment & $92 \pm 5$
\\ \hline 
  \end{tabular}
\end{center}
  \caption{
Collection of results for the excited hyperfine splittings between the
$\psi(2S)$ and $\eta_c(2S)$ from lattice QCD and experiment.
}
\label{tab:hyperfineTwoS}
 \end{table}

\subsection{Stochastic  noise compared to signal}

 We measure the zero momentum disconnected loop $L(t)$ on each
time-slice for  each gauge configuration. This ensemble, for each choice
of operator $\Gamma$  gives us the values of the standard deviation
$\sigma_{\rm obs}$ given in Table~\ref{tb:noise}.  We also,  from our
100 stochastic samples in each case, have the estimate of the  standard
deviation $\sigma_{\rm stoch}$ on the mean of these 100 samples coming
from the stochastic method. We can then deduce the true standard
deviation of the gauge  time slices from $\sigma_{\rm
gauge}=(\sigma^2_{\rm obs}- \sigma^2_{\rm stoch})^{1/2}$. This is
presented in Table~\ref{tb:noise}. Here the normalisation is such that
$M=1+\kappa \dots$.

 \begin{table} 
 \caption{Mesons produced by different operators $\bar{\psi} \Gamma
\psi$. The standard deviation of the loop  operator of eq.
~\ref{eq:loopEXPR}   is presented. Here $\sigma_{\rm stoch}$ is the
error estimated from the 100 stochastic samples used and this is the
used to deconvolute the observed spread to give the true  standard
deviation of the loop ($\sigma_{\rm gauge}$). 
 }
\begin{ruledtabular}
\begin{tabular}{cccccc}
$\kappa$ & $\Gamma$  & $J^{PC}$ &  $\sigma_{\rm obs}$ &  $\sigma_{\rm stoch}$ 
   &$\sigma_{\rm gauge}$ \\
0.135 & $\gamma_5$ & $0^{-+}$ &  33.6 &  13.91  &  30.6  \\ 
0.135 & $\gamma_k$ & $1^{--}$ &  14.7 &  14.45 &  2.7 \\ 
0.135 & $I$   & $0^{++}$      &  53.0 &  15.0  &  50.8  \\ 
0.119 & $\gamma_5$ & $0^{-+}$ &  15.9 &  8.3   &  13.6  \\ 
0.119 & $\gamma_k$ & $1^{--}$ &  9.1  &  9.003 &  1.2  \\ 
0.119 & $I$   & $0^{++}$      &  23.6 &  10.9  &  20.9  \\ 
\hline
\end{tabular}
\end{ruledtabular}
\label{tb:noise}
 \end{table}

 In an ideal world we would have $\sigma_{\rm stoch} \ll \sigma_{\rm
gauge}$  which would imply that no appreciable error arose from the
stochastic methods  employed. The signal in the vector channel is 
dominated by the stochastic error. So for that case, either many more
samples are required or an improved algorithm, such as variance 
reduction~\cite{Thron:1998iy,McNeile:2000xx}. For the other cases, the
stochastic  noise is smaller than the intrinsic gauge fluctuation, and
so more  stochastic samples would not improve the results significantly.

From table~\ref{tb:noise} the error on the heavy-heavy  data is much
less than that from the light-light data. This is a consequence of
increased diagonal dominance of the fermion operator as the mass of the
quark is increased. This is the basis of improved variance
methods~\cite{Thron:1998iy,Wilcox:1999ab,McNeile:2000xx}.

\subsection{Results}

\begin{figure}[th]
\includegraphics[angle=-90,width=\figwidth,clip]{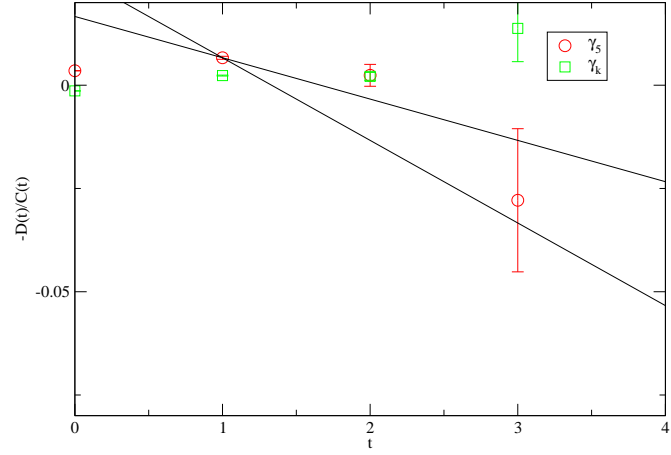}
 \caption{ The ratio of disconnected to connected contributions  as
given by eq.~\ref{dbyc.eq} at the heavy $\kappa$ value of  $0.119$. The
straight lines with slopes  of 0.01 and 0.02 (corresponding to a 
singlet mass 18 and 36 MeV, respectively,  lighter than the non-singlet
in physical units)  are drawn to guide the  eye.
 }
 \label{fg:dbycHeavy}
\end{figure}

\begin{figure}[th]
\includegraphics[angle=-90,width=\figwidth,clip]{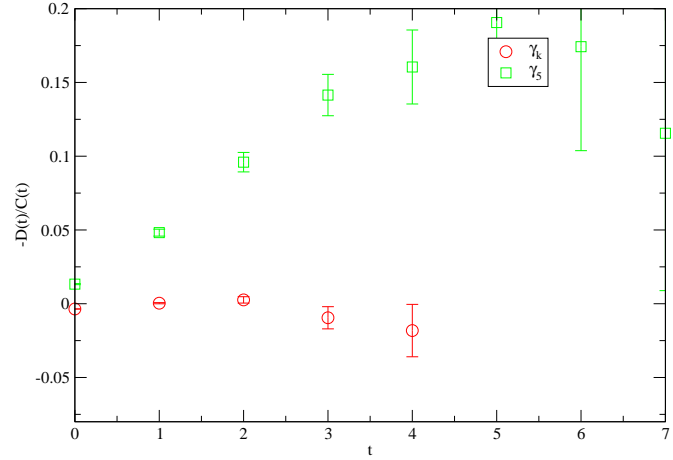}
\caption{The ratio of disconnected to connected contributions 
as given by eq.~\ref{dbyc.eq} at the light $\kappa$ value of 
$0.135$.
}
 \label{fg:dbycLight}
\end{figure}

 We present in figs.~\ref{fg:dbycHeavy} and~\ref{fg:dbycLight} some of
our results for the ratio of the disconnected  correlator to the
connected correlator for the heavy kappa value and light kappa value
respectively. We also measured with fuzzed operators for the
disconnected  diagrams, but they are more noisy than the local operators
we present here.

The error on the  disconnected correlator is much larger than that on
the connected one.  This arises essentially because the absolute error
on the disconnected correlator  stays of the same magnitude as $t$
increases, much as is the case for correlations between  Wilson loops as
used in glueball studies. The connected correlator, in contrast,  has an
approximately constant relative error as $t$ increases.  We are forced
to consider the ratio of disconnected to connected correlator at rather
low $t$-values  because of the  increasing errors on the disconnected
correlator. 

For light (close to the strange quark mass) disconnected contributions
we  find similar results in fig.~\ref{fg:dbycLight} to those obtained
from the lattice previously~\cite{McNeile:2001cr},  namely very little signal
for vector mesons but a large signal for  pseudoscalar mesons that will
increase the singlet  mass over the non-singlet. 
 
\begin{figure}[th]
\includegraphics[angle=-90,width=\figwidth,clip]{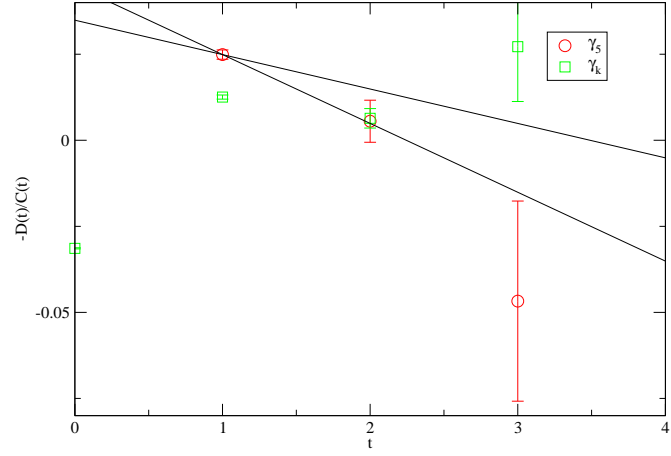}
\caption{The ratio of disconnected to connected contributions 
as given by eq.~\ref{SESAMdbyc.eq} at the heavy $\kappa$ value of 
$0.119$. The SESAM method is used.
The straight lines with slopes 
of 0.01 and 0.02 (18 and 36 MeV in physical units) 
are drawn to guide the 
eye.
}
 \label{fg:dbycHeavySESAM}
\end{figure}

For charm quarks, we present our results   in figures~\ref{fg:dbycHeavy}
 and  \ref{fg:dbycHeavySESAM}. There is only a statistically significant
 signal  at small time values, so no definitive statement can be made.
For the  pseudoscalar, the slope for non-zero $t$ is positive which
corresponds to a  reduction in the singlet mass compared to the
non-singlet. We show  on the figure lines corresponding to a mass shift
of 18 and 36 MeV. This  indicates that we cannot rule out a downward
shift of the $\eta_c$ mass by as much as 36 MeV. For the vector case, the 
signal is smaller (in fig.~\ref{fg:dbycHeavy}) and shows no sign of 
a significant slope.

 As we discuss below, we expect the splittings in the vector channel to 
be small and our results at $t \ne 0$ are consistent with that. We do 
find room, however, for splittings of the order of 20 MeV, particularly 
for the pseudoscalar channel.  We have shown that the contribution of
the singlet correlators to the  hyperfine splitting in charmonium may be
significant. 
From this  calculation, it seems
reasonable that the singlet  correlators could contribute as much as 20
MeV to the  hyperfine splitting. A more definitive estimate requires
more  uncorrelated gauge configurations and/or improved lattice formulations.

\section{Discussion} \label{sect:discuss}

We first recall our previous results from studying the  light singlet
mesons~\cite{McNeile:2001cr} on a lattice.
 The splitting in mass of flavour singlet and non-singlet mesons  with
the same quark content arises from gluonic interactions. The  assumption
that these are small is known as the OZI rule. For the pseudoscalar
mesons  this splitting is not small (it is related to the $\eta$,\
$\eta'$ mass difference), basically because of the impact of the
anomaly. For scalar mesons the splitting is also expected to be large
because of  mixing with the nearby scalar glueball. It is usually
assumed that the  OZI rule is in good shape for the vector and axial
mesons and we found small contributions only.

The picture from the light singlet mesons is that the contribution of
the disconnected piece to the correlators is small unless there is
additional interesting dynamics. In the charmonium system one possible
source of the interesting dynamics is  glueballs. The simplest model is
of a flavour singlet state obtained from the mixing of the flavour
non-singlet state with a glueball, which  causes the states to repel in
energy, often called an  avoided level crossing. We would expect this
mixing to  be strongest when a glueball lay near in energy to the
charmonium state and we  now discuss this.

\begin{figure}[th]
\includegraphics[angle=-90,width=\figwidth,clip]{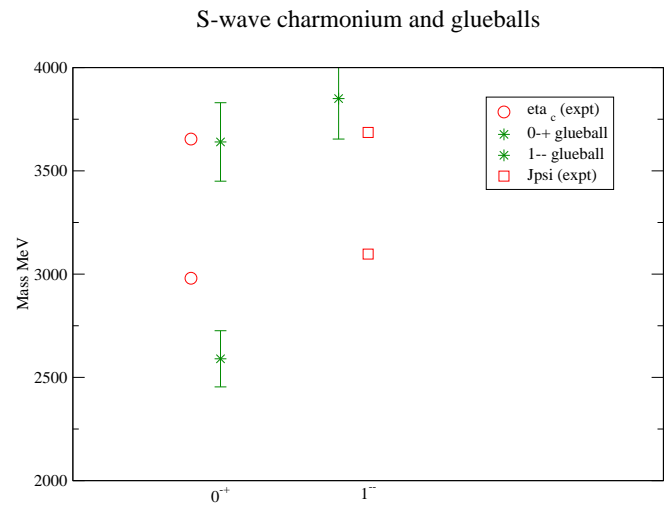}
\caption{
Masses of the experimental 
pseudoscalar and vector states in charmonium  with
the  glueball masses from quenched QCD.
}
 \label{glueball.fg}

\end{figure}

Morningstar and Peardon~\cite{Morningstar:1999rf}  have computed the
excited glueball spectrum in quenched QCD. They obtained masses of
2590(40)(130) MeV and 3640(60)(180) MeV for the ground and first excited
states of the $0^{-+}$ glueball respectively. Morningstar and Peardon
computed the  mass of the $1^{--}$ glueball to be 3850(50)(190) MeV. So
it is not inconceivable that the $\eta_c$ mass (2980 MeV) is effected
more by glueball states than the $J/\psi$ state. In
figure~\ref{glueball.fg} we plot the  masses of the glueballs from
quenched QCD versus the experimental numbers. 
 With these glueball masses, the mixing model  predicts that the singlet
contribution to the  $\eta_c$ will increase the mass, but that the 
singlet contribution to $J/\psi$ will decrease the  mass. This would not
 help to resolve the  discrepancy of the charmonium hyperfine splitting
from non-singlet lattice studies. Moreover this glueball mixing model
gives the opposite sign to the pseudoscalar mass shift than that
indicated by our lattice determination of the disconnected contribution.
Bali has also recently reviewed the influence of glueball states
on the charmonium spectrum~\cite{Bali:2003tp}.

The hyperfine splitting between $\eta_c$(2S) and $\psi(2S)$ states will
also be interesting as it may be affected by the glueball states. The
closeness of the glueball state to the $\psi(2S)$ state has been noticed
by model builders~\cite{Suzuki:2002bz}.  The model for hadron decays for
vector charmonium states involves the emission of three gluons. This
model predicts that the branching ratio for $\psi(2S)\rightarrow\rho\pi$
is much larger than experiment. Attempts have been made  to use the
vector glueball and $\psi(2S)$ mixing to account for this.  If the
glueball states have  large widths, however,  then it is unclear what
the effect of the states will be on the charmonium spectrum.


As well as mixing with glueballs, there are other theoretical models 
which may give guidance on favour singlet mass splittings.
 Isgur and Thacker discuss the origin of the  OZI rule from a quark
model and the large $N_c$ limit of  QCD~\cite{Isgur:2000ts}. Schafer and
Shuryak discuss the OZI rule using instanton-based
methods~\cite{Schafer:2000hn}.


 Another approach is to relate the mass splitting to the fact that the 
decay products (or strongly coupled many-body channels) of the singlet
and non-singlet state are different.  One idea is that  a mass shift
can arise from the energy dependence of the decay width  and will be more
significant for wider resonances.  It will also be possible  that mass
shifts can arise from the back-reaction of the decay  channels to the
effective propagator.    One consequence of this, as has been known for 
a very long time~\cite{cmnstar}, is that the pole in the complex plane
corresponding to  a resonance has an energy whose real part is lower
than the quoted value  which corresponds to a phase shift of $90^0$. 
  This effect of hadron decay on the mass is also an issue for quark
model calculations. Isgur and Geiger~\cite{Geiger:1990yc,Isgur:1999cd}
discuss the effect of decay thresholds on the masses obtained from quark
model calculations. In principle, the effect of hadronic decays can be
introduced into quark models using coupled channel
techniques~\cite{Heikkila:1984wd}. However, it is difficult to write
down a reasonable operator for pair creation.

This issue of the effect of coupled  channels (including open decays) to
the mass of a state is one that can be  illuminated from lattice gauge
theory. This is especially so for singlet states,  since the lattice
enables one to determine the relative contributions  from the connected
and disconnected diagrams separately.

 In the quenched approximation, decays are not treated correctly. This
can be a serious problem: the  connected correlators include only part
of the allowed two-body intermediate states and hence  anomalous results
can be obtained, as for the scalar meson~\cite{Bardeen:2001jm}. Here we
are using  a dynamical quark formalism which is explicitly unitary (at
least in a  world with only $N_f=2$ flavours of quark degenerate in
mass). Within this  formalism we can  add charm quarks without expecting
any significant breakdown of unitarity from neglected charm quark loops
in the vacuum. Then the  correct treatment of charmonium states is to
add the connected and disconnected  contributions, as we have
emphasised. The connected diagram,  once one remembers that light quark
loops are present in the vacuum, contains  intermediate states such as
$D\bar{D}$ and $D^* \bar{D}$ etc.  It does not contain  charmless
intermediate states.  The hadronic decays of those charmonium  states
below the $DD$ threshold are necessarily to charmless intermediate
states. These  are just the charmless states that are allowed as
intermediate states in the disconnected diagram evaluated on the
lattice.  Thus there is a link between  the disconnected  diagram  and
the hadronic decay of the charmonium state.  For an OZI-violating decay,
the charm quark and anti-quark must annihilate which is similar process
to the contribution of the disconnected diagrams
(figure~\ref{fg:disLOOPS})  to the singlet correlator. This link is not
unambiguous for light quarks: for example the substantial $\eta$ -$\pi$ 
splitting (in a world with $N_f=2$) arises from the disconnected diagram
 but no hadronic decay of the $\eta$ is allowed energetically.  There
does not seem to be a simple quantitative  link  between the mass shift
caused by the disconnected loop in the correlator and the width of the
state.

 This link is explicit in a perturbative treatment of charmonium (even
more so for $\bar{b} b$): one can evaluate the OZI violating
contributions to  charmonium from multiple gluon exchange. The
pseudoscalar meson allows  two gluon exchange and so should have much
larger effects than for the  vector meson where three gluons are needed.
Moreover the hadronic decays are to multi light-quark states  created
from these two (or three) gluon intermediate states.

The computation of strong decay widths from lattice QCD is a hard
problem. There are formalisms available, but the numerical calculations
are quite difficult.  Some of the issues about decay widths and lattice
QCD have been recently been
reviewed~\cite{McNeile:2003dy,Michael:2003vw}.  In a large lattice
spatial volume,  the effect of coupled  two-body decay channels on the
mass of a state  is already taken account of by the formulation,
provided one uses a unitary  theory with the same valence quarks as
sea
quarks.
Thus one should not expect any shift from decay channels. One
example of this is that the baryon decuplet shows experimentally an
equal mass spacing arising from the number of strange quarks
present, even though the widths of the members vary from 120 MeV
($\Delta$) to  stable ($\Omega$).

On a lattice, at smaller volumes, the two-body momentum states become
discrete  and this induces small shifts in mass. These have been exploited
by L\"uscher to yield information  about two-body scattering from the
lattice.  An example of a shift in the $\rho$ mass on a lattice from its 
coupling to $\pi \pi$ has also been studied~\cite{McNeile:2002fh}.

For hadrons containing only light quarks it is  difficult to compute
decay widths from first principles on a lattice.  For charmonium, decay
widths can in principle be computed using the NRQCD factorisation
formalism~\cite{Bodwin:1995jh,Bodwin:2002hg}, or from older techniques
based on factorising the decay width into a perturbative part and
the wave function at the origin~\cite{Kwong:1988ak}.

 Thus we should interpret our results as giving an indication of the
strength and  sign of OZI violating contributions to the heavy meson
spectrum. These need not correspond to those observed experimentally 
because we would need to extrapolate our lattice results to the
continuum limit and to more realistic  quark masses (including a third
flavour). This extrapolation  in quark mass could be quite delicate,
because of issues such as  mixing and decays, as discussed above. We
note that $\bar{b}c$ mesons will not have these singlet contributions
and so the hyperfine splitting for them should agree with  a lattice
calculation using only connected contributions. This  may  be a useful
experimental source of input into the composition of such states.

Both the $\eta_c$ and $J/\psi$ are below the threshold for 
$D\overline{D}$ decays. However OZI-violating hadronic  decays are
allowed. The current summary of the $\eta_c$ properties in the particle
data table~\cite{Hagiwara:2002fs}, quotes the width of the $\eta_{c}$ as
$16^{+3.8}_{-2.1}$ MeV. However, the latest results for the  width of
the $\eta_c$ are larger than the number in the particle data table.
CLEO~\cite{Brandenburg:2000ry}, BES~\cite{Bai:2003et}, and
BaBar~\cite{Wagner:2003qb} obtain for the $\eta_c$ width: $27 \pm 5.8
\pm 1.4$ MeV, and $17 \pm 3.7 \pm 7.4$ MeV, and  $33 \pm 2.5$ MeV
respectively. CLEO~\cite{Brandenburg:2000ry} note that a larger width
($\sim 28$ MeV) for  $\eta_c$ improves agreement with experiment for the
 next to leading order perturbative QCD expressions~\cite{Kwong:1988ak} 
for  the ratio of  the decay width to the two photon width  (which is
more precisely known). This agreement with perturbative estimates also
suggests that mixing with glueballs is not the dominant mechanism for the 
$\eta_c$ decay, and hence for the $\eta_c$ mass contribution from 
disconnected diagrams.
 We note that an $\eta_c$  width of 30 MeV is comparable to the  width
of some $\overline{c}c$ mesons above the $\overline{D}D$ threshold, such
as the $\psi(3770)$ with a width of 24 MeV, although typically the
widths of $\overline{c}c$ mesons above threshold are above 80 MeV.

The width of the $J/\psi$ is accurately known and is 87
keV~\cite{Hagiwara:2002fs}. The reason for the smaller  width of the
vector mesons is that the leading order perturbative corrections are
$O(\alpha_2^3)$ for the  vector channel, but $O(\alpha_2^3)$ for the 
pseudoscalar channel.

Although a width of 30 MeV is small relative to the mass of the
$\eta_{c}$, this width is not small relative to the hyperfine
splitting.  It is the mass splittings that are the significant
quantities in charmonium.
So it is not unreasonable that there is a shift coming from
OZI-violating intermediate states in the mass of the $\eta_{c}$ of the
order of 20 MeV and that the mass of the $J/\psi$ is unaffected. This
can be  substantiated by more accurate lattice evaluations, our
calculation leaves room for an  effect of such a magnitude but with large
errors.

We have shown that it is necessary to compute the disconnected
contributions to the charm correlators to obtain accuracies under 3 MeV
for mass splittings~\cite{Davies:2003ik}. 
Studies with anisotropic lattices may be useful to
sample more intermediate points, but a long reach in physical time is
also required. The study of disconnected charm quark loops may also be
useful for looking at hidden charmonium~\cite{Brodsky:2001yt}.

\section{Acknowledgements}

We thank Alex Dougall and Chris Maynard for discussions.
The lattice data was generated on the Cray T3D and T3E systems at EPCC
supported by, EPSRC grant GR/K41663, PPARC grants GR/L22744 and 
PPA/G/S/1998/00777. We are grateful to the ULgrid project of the 
University of Liverpool for computer time.
The authors acknowledge support from EU 
grant HPRN-CT-2000-00145 Hadrons/LatticeQCD.




\end{document}